\documentstyle[12pt]{article}

\begin{document}
\input epsf.tex
\rightline{McGill/96-19}
\vspace{.5 cm}
\begin{center}
{\Large\bf Multiple Reggeon Exchange from Summing QCD }\\
\vspace {.3cm}
{\Large\bf Feynman Diagrams}\\
\vspace{.5 cm}
{Y.J. Feng$^{\dag}$ and C. S. Lam$^*$}\\
\bigskip
{\it Department of Physics, McGill University,\\
3600 University St., Montreal, P.Q., Canada H3A 2T8}
\end{center}

\begin{abstract}
Multiple reggeon exchange supplies subleading logs that may be used to restore
unitarity to the Low-Nussinov Pomeron, provided it can be proven that the sum
of Feynman diagrams to all orders gives rise to such multiple regge exchanges. 
This question cannot be easily tackled in the usual way except for very 
low-order diagrams, on account of delicate cancellations present in the sum
which necessitate individual Feynman diagrams to be computed to
subleading orders. Moreover, it is not clear that sums of
high-order Feynman diagrams with complicated criss-crossing of lines can
lead to factorization implied by the multi-regge scenario. Both of these
difficulties can be overcome by using the recently developed nonabelian 
cut diagrams.
We are then able to show
that the sum of $s$-channel-ladder diagrams to all orders does lead to such
multiple reggeon exchanges.   
\end{abstract}

\section{Introduction}
\setcounter{equation}{0}
The gluon in QCD reggeizes in the leading-log approximation. 
The coupling constant
($g$) and the energy ($\sqrt{s}$) of sum of one-reggeized-gluon
(1rg) diagrams come in the form $g^2(g^2\ln s)^p$, where $g^2\ll 1$,
$g^2\ln s=O(1)$, and $p$ is a non-negative integer. The sum of 2rg diagrams
are of the form $g^4(g^2\ln s)^p$,
and more generally the $m$rg amplitude is given by sums of terms of the form
$g^{2m}(g^2\ln s)^p$.

For quark-quark elastic scattering at high energy and fixed momentum transfer,
the color exchanged in a 1rg amplitude is an octet, and that for a 2rg
amplitude is either an octet or a singlet. The 2rg amplitude being a factor
$g^2\ll 1$ down from the 1rg amplitude, its octet contribution can be
neglected, but its singlet part must be kept, for there is no
competing contribution from the 1rg amplitude. This singlet part is just
the Pomeron proposed by Low and Nussinov \cite{FE} .

The leading-log Pomeron amplitude obtained this way \cite{LN,CY}
violates unitarity. It
leads to a total cross section with a power growth in $s$,
which is forbidden by the Froissart bound. To unitarize the BFKL equation \cite{LN} 
it is therefore necessary to include subleading-log contributions.

Subleading logs are notoriously difficult to extract from Feynman diagrams.
One may try to compute some low-order diagrams to get information \cite{FA} but
it is almost certain that this cannot be carried out to all orders.
However, subleading-log contributions to the Pomeron do not necessarily
require subleading logs from sums of Feynman diagrams. For example,
leading-log calculation of Feynman diagrams contributing to 2rg exchanges
gives subleading-log
correction to the octet amplitude. Hence there is a hope in unitarizing
the Pomeron amplitude without having to invoke difficult
subleading-log calculations.
In fact, $s$-channel unitarity in all color channels is formally satisfied
if all multiple rg exchanges are added in, as shown in the
{\it reggeized diagrams} in Fig.~1.
To allow for shadows produced by
inelastic scatterings, we must include production of gluons from the rg's
in the intermediate states, though creation of quark pairs will be ignored
in the present discussion.
Whether this proposal of unitarization \cite{CY}, without including
subleading terms in individual sums of Feynman diagrams, is
correct or not remains an open question which we simply cannot discuss
until more is known.

\begin{figure}
\vskip -0 cm
\centerline{\epsfxsize 3 truein \epsfbox {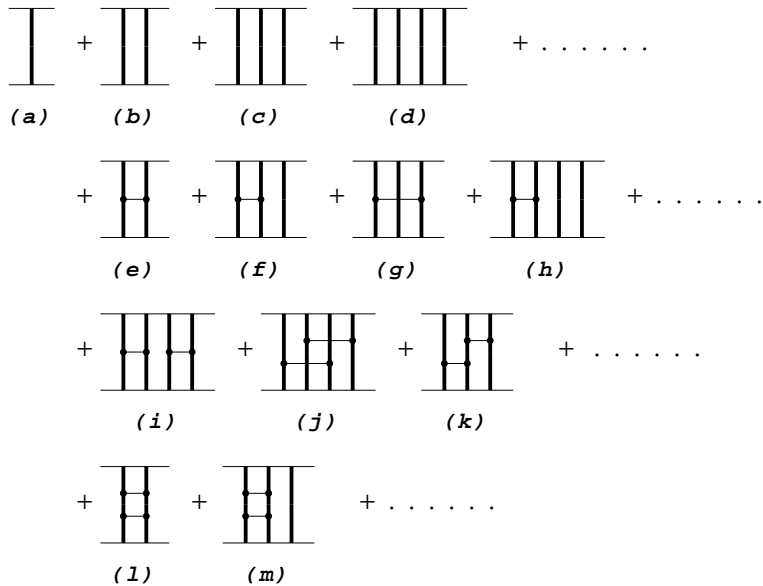}}
\nobreak
\vskip -2 cm\nobreak
\vskip .1 cm
\caption{Multi-reggeon exchange diagrams.}
\end{figure}

The necessity of including multiple-rg exchange diagrams can be understood in
a completely different way, totally within the framework of
leading-log approximations.
Imagine we are dealing with an $SU(N_c)$
color theory in which quarks carry an arbitrary color.  Then there are many
independent color
amplitudes, more so if $N_c\gg 1$ and
quarks carrying a large color. To retain the {\it leading-log} contribution
of every one of these independent color amplitudes, we must retain the $m$rg
contributions for every $m$. So even staying within leading logs, those
multiple-rg exchanges are required for
a color $SU(N_c)$ theory with arbitrary quark colors.
For that reason we shall carry out our calculations below
for an arbitrary $N_c$ and every quark color. Since spin is unimportant
in high energy scattering \cite{LN,CY}, this has the added advantage that whatever we obtain
is automatically valid for gluon-gluon scattering as well.

What is missing in this scheme is
the proof that the {\it reggeized factorization
hypothesis}
is indeed correct, that the sum of Feynman diagrams in the leading-log
approximation does factorize into sums of these multiple-rg amplitudes.

To be sure, the hypothesis has been verified explicitly up to the
6th order, and partially up to the 8th and 10th orders \cite{LN,HC,NT},
but because of the presence of delicate cancellations, it
is difficult to carry similar calculations to higher orders. In fact,
these delicate cancellations have not been completely verified even in the 8th
and the 10th orders.

The problem is the following. In Feynman gauge calculation
which we shall adopt throughout, the leading-log contributions in some color
amplitudes get cancelled out when several Feynman diagrams are summed 
\cite{HC,NT,YJ}.
Consequently, to compute the {\it sum} of diagrams to leading-log accuracy,
we need to calculate individual diagrams
to subleading-log precision.
Occasionally this can be accomplished
without much pain by using $s\leftrightarrow  u$ symmetry, but more often not.
To the
extent that subleading logs are very difficult to compute,
calculations to higher-order diagrams
do appear to be quite forbidding.

Even if we manage to overcome this hurdle, the verification of
reggeized {\it factorization} from sums of Feynman diagrams, with lines
crossing one another in very complicated ways, would still seem to be
extremely difficult.

Fortunately there is a chance to overcome
both of these difficulties by using
{\it nonabelian cut diagrams} \cite{YJ} in place of the usual Feynman diagrams.
Nonabelian cut diagrams
are resummations of Feynman diagrams with these delicate
cancellations removed, so that each of them can be
computed just in
the leading-log approximation. Moreover, factorization is natural to
the nonabelian cut diagrams, because they are
the graphical manifestation of a `multiple commutator
formula', which in turn was derived from a `factorization formula' \cite{CS},
and it is this same factorization formula that will be used to demonstrate
reggeized {\it factorization} hypothesis for a class of diagrams.

Nonabelian cut diagrams will be reviewed in the next section. Some of their
properties, including the assertion of the absence of delicate cancellations
in these cut diagrams but their presence in usual Feynman diagrams,
will be discussed in Sec.~3. In this paper we shall study in detail,
and be able to prove,
the reggeized factorization hypothesis for a particular simple class
of diagrams, the $s$-channel-ladder diagrams. It is this class of diagrams
in QED that can be summed up into an explicit eikonal form, so one would
expect
it to be the simplest set to study in QCD as well
for multiple-rg exchanges and
unitarization.
However, with color complication, the QCD case is
much harder to deal with than the QED case, the details of which are
discussed in Sec.~4.
For more complicated nonabelian cut diagrams in QCD, we are not yet
able to prove the reggeized factorization, but the success for the
$s$-channel-ladder diagrams is encouraging.
Finally, a
short summary and outlook are provided in Sec.~5.

\section{Nonabelian cut diagrams}
\setcounter{equation}{0}

Nonabelian cut diagrams \cite{YJ} represent a resummation of Feynman diagrams.
They are not the same as Cutkosky cut diagrams which compute
discontinuities of single Feynman diagrams.

A nonabelian cut diagram differs from a Feynman diagram
in having certain `high speed'
propagators cut. The cut lines occur among  those
carrying a large momentum $p$, with comparatively
small amount of momenta $q_i$ transferred away at each interaction vertex.
In that case,
the approximation
\begin{eqnarray}
(p+\sum_{j=1}^iq_j)^2-m^2\simeq 2p\cdot \sum_{j=1}^iq_j
\equiv \sum_{j=1}^i\omega_j
\end{eqnarray}
is valid, so the denominators of the Feynman
propagators for these lines are simplified to
$(\sum_{j=1}^i\omega_j+i\epsilon)^{-1}$.

\begin{figure}
\vskip -0 cm
\centerline{\epsfxsize 3 truein \epsfbox {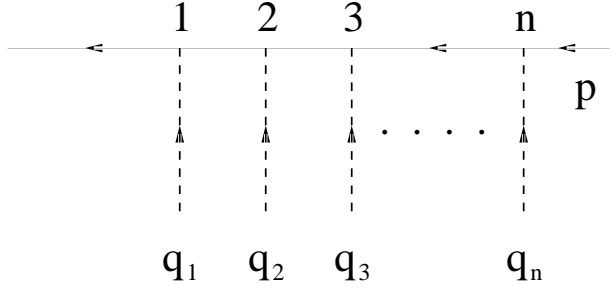}}
\nobreak
\vskip -2 cm\nobreak
\vskip .1 cm
\caption{A tree diagram with $n$ bosons emitted or absorbed.}
\end{figure}

Within this approximation, the
QCD tree diagram for a propagating quark shown in Fig.~2 is
\begin{eqnarray}
-2\pi i \delta(\sum_{j=1}^{n}\omega_j)\left(\prod_{i=1}^{n-1}{1\over \sum_{j=1}^i\omega_j+i\epsilon}\right)\cdot t_1t_2\cdots t_n\cdot V 
\equiv a[12\cdots n]\cdot t[12\cdots n]\cdot V\ ,
\end{eqnarray}
where $t_i$ are the color matrices of the quark, and
$t[12\cdots n]\equiv t_1t_2\cdots t_n$. The numerator with the normalization convention
$\bar u u=1$ can be approximated by
\begin{eqnarray}
V={1\over 2M}\prod_{i=1}^n(2p^{\mu_i})\ ,
\end{eqnarray}
where $\mu_i$ are the Lorentz indices of the gluons and $M$ is the quark mass.

The tree diagram in
Fig.~2 will be denoted by $[12\cdots n]$, according to the order of the gluons. If
the gluons
are labelled differently, say as $[\sigma_1\sigma_2\cdots \sigma_n]$,
then the corresponding spacetime amplitude and color factor will be similarly
designated
as $a[\sigma_1\sigma_2\cdots \sigma_n]$ and $t[\sigma_1\sigma_2\cdots \sigma_n]$.

Before discussing the nonabelian cut diagrams it is necessary to introduce
some notations. If $[T_i]$ are tree diagrams, then $[T_1T_2\cdots T_A]$ represents
the tree diagram obtained by merging these $A$ trees. For example, if
$[T_1]=[123]$
and $[T_2]=[45]$, then $[T_1T_2]=[12345]$. The notation $\{T_1;T_2;\cdots ;T_A\}$,
on the other hand,
is used to denote the {\it set} of all tree diagrams obtained by
{\it interleaving} the trees $T_1, T_2, \cdots , T_A$ in all possible ways.
This set contains
$(\sum_a n_a)!/\prod_an_a!$ trees if $n_a$ is the number of gluon lines in the
tree $T_a$. In the example above, $\{T_1;T_2\}$
contains the following $5!/3!2!=10$ trees: [12345], [12435], [12453], [14235],
[14253],
[14523], [41235], [41253], [41523], and [45123].

Correspondingly,
$a\{T_1;T_2;\cdots ;
T_A\}$ will represent the sum of the amplitudes $a[T]$ for every tree $T$
in this set.

The nonabelian cut diagram \cite{YJ} is derived from the
the {\it multiple commutator formula} \cite{CS}, which states that
\begin{eqnarray}
\sum_{\sigma\in S_n}a[\sigma]t[\sigma]=\sum_{\sigma\in S_n}a[\sigma]_c
t[\sigma]'_c\ .
\end{eqnarray}
This is a resummation formula for the nonabelian tree amplitude (2.2),
summed over all $n!$ permutations
$[\sigma]\equiv[\sigma_1\sigma_2\cdots \sigma_n]$ of $[12\cdots n]$. The spacetime
part of the {\it cut
amplitude} $a[\sigma]_c$ is obtained from the {\it cut diagram} $[\sigma]_c$,
and the color factor $t[\sigma]'_c$
is obtained from the {\it complementary cut diagram} $[\sigma]'_c$.
All of these will be explained below.

The multiple commutator formula in turn was derived from the
{\it factorization formula} \cite{CS}, which states that
\begin{eqnarray}
a\{T_1;T_2;\cdots ;T_A\}=\prod_{a=1}^Aa[T_a]\ .
\end{eqnarray}
This is a sum rule expressing factorization of sums of certain tree amplitudes.
It is this same formula that proves to be invaluable
in showing the reggeized {\it factorization} later.

A special case of the factorization formula is well known.
If $n_a=1$ for every $a$ so that the tree $[T_a]=[a]$ is simply a vertex,
then $\{1;2;\cdots ;A\}$ is the set of $A!$ permutation of the
tree $[12\cdots A]$, and the factorization
formula is just the well-known {\it eikonal formula} \cite{HCT}.

We shall now proceed to define the cut diagrams and the cut amplitudes.
To each Feynman tree diagram $[\sigma]=[\sigma_1\sigma_2\cdots \sigma_n]$ of the
type shown in Fig.~2, we associate with it
a {\it cut diagram} $[\sigma]_c$ by putting cuts on specific fermion
propagators as follows. Proceed from left to
right, put a cut after a gluon if and only if a smaller number does not
occur to its right. Continuing thus until reaching the end of the tree,
and we get the cut diagram. An external line would be considered equivalent
to a cut so there is never an explicit cut put in at the end of the tree.

The written notation for a cut will be a vertical bar behind a gluon.
Using that notation,
here are some illustrations of where cuts are put into Feynman trees:
$[1234]_c=[1|2|3|4]$, $[3241]_c=[3241]$, and $[2134]_c=[21|3|4]$.

The {\it complementary cut
diagram} $[\sigma]'_c$ is one where lines cut in $[\sigma]_c$ are not cut in
$[\sigma]'_c$, and vice versa. Thus $[1234]'_c=[1234]$,
$[3241]'_c=[3|2|4|1]$, and $[2134]'_c=[2|134]$.

The spacetime part of the cut
amplitude, $a[\sigma]_c$,  is simply the Feynman amplitude $a[\sigma]$
with the cut propagator taken to be
$-2\pi i\delta(\sum_j\omega_j)$
instead of the usual $(\sum_j\omega_j+i\epsilon)^{-1}$. In this way it is
the same cut propagator as in the Cutkosky cut diagram, but here
cuts are placed only on
high speed fermion lines, and as (2.4) indicates, the nonabelian cut diagrams
represent a resummation and not a discontinuity.

The color factor
$t[\sigma]'_c$ is determined from the complementary cut diagram $[\sigma]'_c$.
It is obtained from $t[\sigma]$ by replacing the product of
color matrices separated by cuts with their commutators. For example,
$t[1234]'_c=t[1234]=t_1t_2t_3t_4$,
$t[3214]'_c=t[3|2|4|1]=[t_3,[t_2,[t_4,t_1]]]$,
and $t[2134]'_c=t[2|134]=[t_2,t_1]t_3t_4$.

A Feynman diagram for quark-quark scattering can be obtained by connecting
two trees like Fig.~2 together via the gluon lines, perhaps with the help of
triple gluon and four gluon vertices and other propagators in between.
Since (2.4) is valid for offshell gluons, it can be applied to one of the
two quark trees carrying large momentum even though it is tied up
in a loop diagram. Unless otherwise
stated, relation (2.4) will always be applied to the upper quark tree, so cuts
are normally made only on this line. The rest of the propagators remain uncut
and the diagram is otherwise the same as an ordinary Feynman diagram.

\section{Color factors}
\setcounter{equation}{0}

We define {\it regge color factors} to be the color factors appearing in
the reggeized diagrams of Fig.~1. If the reggeized factorization
hypothesis is correct,
only regge color factors can be present when all Feynman diagrams are summed.

We define {\it primitive color factors} as regge color factors that cannot
be pulled apart into smaller
units by sliding along the pair of quark lines. In other words, 
they are those that
remain connected after the upper and lower quark lines are removed.
Thus 1(a), 1(e), 1(k), 1(l)
are primitive, but none of the others in Fig.~1 are.

It is shown in App.~A that in the leading-log approximation, every regge
color factor that has the same number of primitive color units can
be considered to be the same, irrespective of how the units are placed.
Thus 1(f) is the same as 1(g), and 1(i) is the same as 1(j).
The number of times $f_\alpha$ a primitive color factor $\phi_\alpha$
occurs completely specifies a regge color factor, which will
be denoted by $\Phi=\prod_\alpha \phi_\alpha^{f_\alpha}$.

The primitive color factor in 1(a) will be designated as
I. The primitive color factor for 1(e) will be designated as
H. The regge color factors of 1(b), 1(c), 1(d), 1(f),
1(g), 1(h), 1(i), 1(j) are then $I^2$, $I^3$, $I^4$, $HI$, $HI$, $HI^2$, $H^2$,
$H^2$. In the notation of Refs.~[5,7],
$I={\bf G_1}, H={\bf G_3}, I^2={\bf G_2}$, and $I^3={\bf G_4}$.

The color factor of an ordinary Feynman diagram can be decomposed into
sums of regge color factors \cite{CY,HC} by using the commutation relation
\begin{eqnarray}
[t_a,t_b]=if_{abc}t_c\ ,
\end{eqnarray}
as well as the sum rules
\begin{eqnarray}
f_{abc}f_{abd}=2c\delta_{cd}\ ,\quad
i^3f_{adg}f_{bed}f_{cge}=cif_{abc}\ ,
\end{eqnarray}
where $c=N_c/2$ for a color
$SU(N_c)$ group, and $t_a$ is the color matrix of the quark in any
representation. In particular, if it is in the adjoint representation, then
$(t_a)_{bc}=if_{bac}$, and this is represented graphically by a triple-gluon
vertex read in clockwise order.
These relations are shown in Fig.~3
where a cut represents a commutator.
Using these figures,
decomposition can be accomplished in a graphical way.
For details and concrete illustrations, see Ref.~[5].
\begin{figure}
\vskip -0 cm
\centerline{\epsfxsize 3 truein \epsfbox {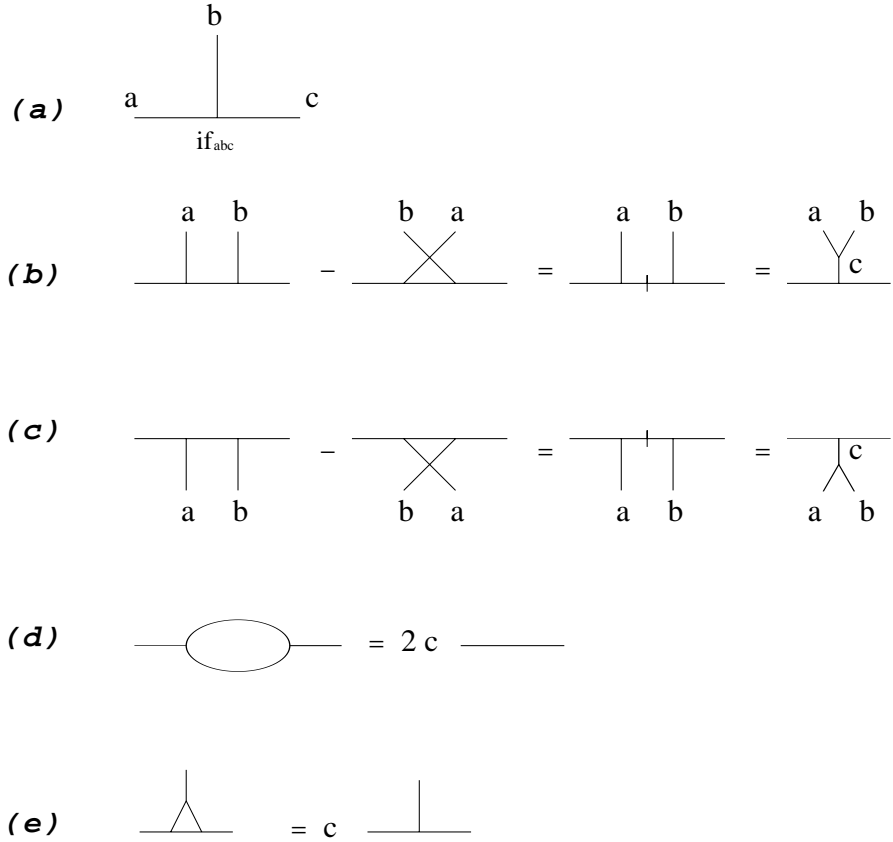}}
\nobreak
\vskip -2 cm\nobreak
\vskip .1 cm
\caption{Color matrices and their relations (3.1) and (3.2) in graphical
forms.} 
\end{figure}

Decomposition of color factor of a cut diagram into regge color
factors can be carried out in a similar way, also graphically \cite{YJ}. 

A complementary cut diagram
with $m-1$ {\it uncut} propagators on the upper quark line contains
only regge color factors
with {\it at most} $m$rg exchanges. This statement is a simple consequence of
the graphical construction procedure for the regge color factors \cite{YJ}.
See App.~A.

With this simple property, we can now understand why delicate cancellations
do not occur in {\it nonabelian cut diagrams}. To see that,
it is sufficient to show that the spacetime amplitude for an $m$rg color factor
does not have more $\ln s$ power than
$g^{2m}(g^2\ln s)^p$, thus there is no need to cancel out higher powers of
$\ln s$ to get to the regge behavior. We shall also see that this is not
generally so in {\it Feynman diagrams}, hence delicate cancellations are
necessary there.

Consider a complementary cut diagram with $m-1$ uncut lines. This contains
regge color factors with at most $m$rg exchanges. The corresponding
cut diagram for the spacetime amplitude has $m-1$ cut lines. Now each loop
in a spacetime diagram can contribute at most one $\ln s$ factor, but
this factor will be absent in any loop containing a cut propagator. This is so
because the Feynman propagator giving rise to the $\ln s$ factor through
integration is now replaced by a $\delta$-function \cite{HC,YJ}.
With $m-1$ cut propagators, $m-1$ potential $\ln s$ factors are lost, so
the spacetime amplitude can grow as most
like $g^{2m}(g^2\ln s)^p$, as claimed.

If these inequalities are saturated, that a spacetime cut diagram with $m-1$
cuts grows like $g^{2m}(g^2\ln s)^p$, then the
cut diagram is said to be {\it saturated}. Only saturated diagrams are needed
in leading-log computations; unsaturated ones are negligible in comparison.

This also means that in the leading-log approximation, there is no need
to include in a complementary cut diagram with $m-1$ uncut lines those
regge color factors with less than $m$rg exchanges.

From these discussions we can also see why delicate cancellations are generally
expected for Feynman diagrams if the reggeization hypothesis is valid.
A Feynman diagram has no cut in its spacetime diagram, nor its color factor.
The former tends to give rise to more $\ln s$ factors than a corresponding
cut diagram,
and the latter will generally yield regge color factors with
larger $m$. For both reasons there are too many $\ln s$ powers compared to
the reggeized behavior of $g^{2m}(g^2\ln s)^p$, so delicate cancellations
eliminating these powers must take place.

\section{$s$-channel-ladder cut diagrams}
\setcounter{equation}{0}
\subsection{Description of the diagrams}
$s$-channel-ladder diagrams are obtained by joining together the gluons of two
quark trees like Fig.~2. If we number the gluons attached
to the lower quark tree in the order $[123\cdots n]$, then
the order of gluons along the upper quark tree can be
used to specify the whole $s$-channel-ladder diagram.
Cut diagrams are determined by the rules discussed in Sec.~2, and cut
propagators on the upper quark tree
will be indicated by a vertical bar as before.
In this notation, Figs.~4(a) to 4(k) are respectively
$[1|2|3|4], [21|3|4], [31|2|4], [1|32|4]$, $[41|2|3], [1|42|3], [1|2|43],
[21|43], [31|42],$ $[41|32], [321|4]$.
Unless otherwise stated, propagators along the lower quark tree will remain
uncut, so when we refer to cut and uncut propagators, we will usually be
speaking of those along the upper quark tree.

We will use the abbreviation SC to denote {\it $s$-channel-ladder cut diagram},
and the notation SCC to denote {\it $s$-channel-ladder complementary cut
diagram}. The former is used to compute spacetime amplitudes, and the latter
used to compute color factors. Cut propagators in SC  become uncut propagators
in SCC and vice versa.

According to the discussion in Sec.~3,  in the leading-log approximation,
we need to retain only the $m$rg regge color factors from a SCC diagram
with $m-1$ uncut propagators, and only the saturated
SC diagrams with $m-1$ cuts on the
upper tree, whose  $\ln s$ power is given by $g^{2m}(g^2\ln s)^p$.
It is shown in App.~B that the saturated SC diagrams are those
without adjacent uncut propagators on the upper tree. For example,
Figs.~4(a) to 4(j) are  saturated, but 4(k) is not.
\begin{figure}
\vskip -0 cm
\centerline{\epsfxsize 3 truein \epsfbox {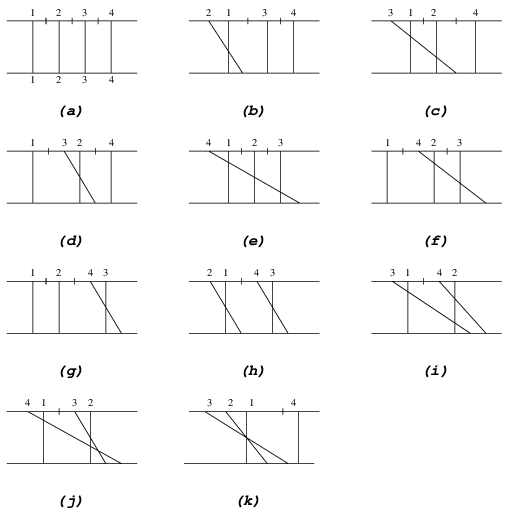}}
\nobreak
\vskip .3 cm\nobreak
\vskip .1 cm
\caption{Examples of 8th order $s$-channel-ladder cut diagrams.}
\end{figure}

From the rules of Sec.~2, we conclude that
all propagators of the planar diagram are to be cut, as in $[1|2|\cdots |n]$. For
other
SC diagrams, cuts are placed behind a number if and only if there is not a
smaller number to its right.

We can phrase this in a way independent of the numberings of the gluon lines.
All SC diagrams are obtained from the planar diagram
by pulling the upper ends of some gluon lines leftward.
Once a gluon line is moved, the cut to its right also disappears.

We shall always draw the gluon lines in the {\it planar} diagram vertical, as
in 4(a). Any other SC diagram will consist of a number of slanted lines
mixed in with the vertical lines. The propagator to the right of a slanted
line is uncut, but every other propagator is cut. See Fig.~4 for illustrations.
The upper ends of two slanted lines in a {\it saturated} SC diagram
must not be adjacent.
In other words, each slanted line will find a vertical line to its right along
the upper tree, forming together a {\it skeleton cross}.
These two lines will be called the {\it skeleton lines}, and the rest of the
vertical lines will be called the
{\it mobile lines}. Mobile lines have cuts on both sides, or a cut on one side
and an external line on the other, while skeleton lines have a cut propagator
or an external line on one side, and an uncut propagator on the other. The
propagator inside the skeleton cross of a saturated SC is always uncut.

It is the opposite in a saturated SCC. The propagators inside the skeleton
crosses are cut but other propagators along the upper quark line remain uncut.

Vertical skeleton lines will be called {\it $v$-lines}, slanted skeleton
lines will be called {\it $s$-lines}, and mobile lines will be called
{\it $m$-lines}. We shall label  the $v,s,m$ lines of a diagram respectively by
$v_i,s_i\ (1\le i\le k)$, and $m_j\ (1\le j\le b=n-2k)$.
$n$ is the total number of gluon lines and
$k$ is the number of skeleton crosses in the diagram.

In terms of these labels,
the lower tree of a SC or SCC diagram is a tree in the set
\begin{eqnarray}
{\cal S}'_{k,b}\equiv\{v_1s_1;v_2s_2;\cdots ;v_ks_k;m_1;m_2;\cdots ;m_b\}\ .
\end{eqnarray}
See Sec.~2 for notations.
Conversely, by erecting vertical lines over $v_i$ and $m_j$, and slanted
lines over $s_i$ with upper ends lying just to the left of $v_i$, we can
reconstruct an SC or SCC diagram from any tree in (4.1). However,
any permutation of the $k$ skeleton crosses and the $b$ mobile lines
will give the same diagram, so the number of distinct   diagrams is
only $(2k+b)!/2^kk!b!$. We shall denote the set of {\it distinct} diagrams
by ${\cal S}_{k,b}$. Its trees will be taken from those in ${\cal S}'_{k,b}$
with $v_1<v_2<\cdots <v_k$, and $m_1<m_2<\cdots <m_b$, where $a<b$ means line
$a$ is to the left of line $b$ at the lower end.
We shall refer to trees with these restrictions
as {\it trees in ascending order}.

If we need not know $k$ and $b$
explicitly, we shall simply write the sets as ${\cal S}'$ and ${\cal S}$ respectively.

\subsection{Color factors}
Given a saturated SCC diagram $[T]\in{\cal S}$,
we want to know how to decompose its color factor into
combinations of regge color factors, and conversely given a regge
color factor, how to find all the $[T]\in{\cal S}$ containing it.

Decomposition is carried out in the following way. The graphical
commutation relations in Fig.~3 is used to pull the lower end of the $s$-lines
leftward,
until it sits just to the right of the corresponding
$v$-line, a position which we shall refer to as the
{\it home position}. As a result, The SCC diagram is given by a sum of many
{\it reduced diagrams},
each of which having the bottom end of the $s$-lines lying to the left of their
original positions, unless of course they were already in their home positions
to begin
with. Every $s$-line not at its home position should have a cut to its
right along the lower tree; an $s$-line at the home position may or may not
have such a cut; both are allowed.
The sum of all such combinations, weighted with a
minus
sign iff there is an odd number of cuts, is the decomposition desired.
A simple illustration is given in Fig.~5.
\begin{figure}
\vskip -0 cm
\centerline{\epsfxsize 3 truein \epsfbox {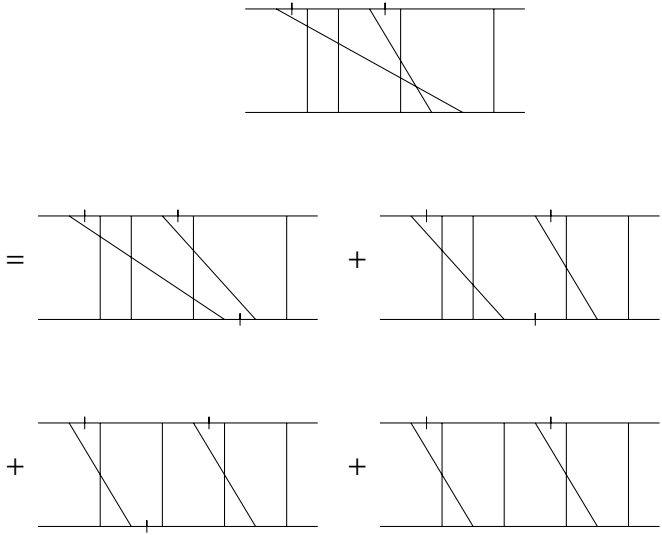}}
\nobreak
\vskip -2 cm\nobreak
\vskip .1 cm
\caption{An example of the decomposition of SCC diagrams into sums of
reduced diagrams.}
\end{figure}

A reduced diagram differs from an SCC diagram in that it is represented
by a lower tree with {\it cuts}. As discussed above, cuts
always occur to the right of $s$-lines, unless they are already
in their home positions.
We shall use the symbol ${\cal C}$ to denote the set of {\it distinct} reduced
diagrams, represented by lower cut trees in ascending order.
\begin{figure}
\vskip -0 cm
\centerline{\epsfxsize 3 truein \epsfbox {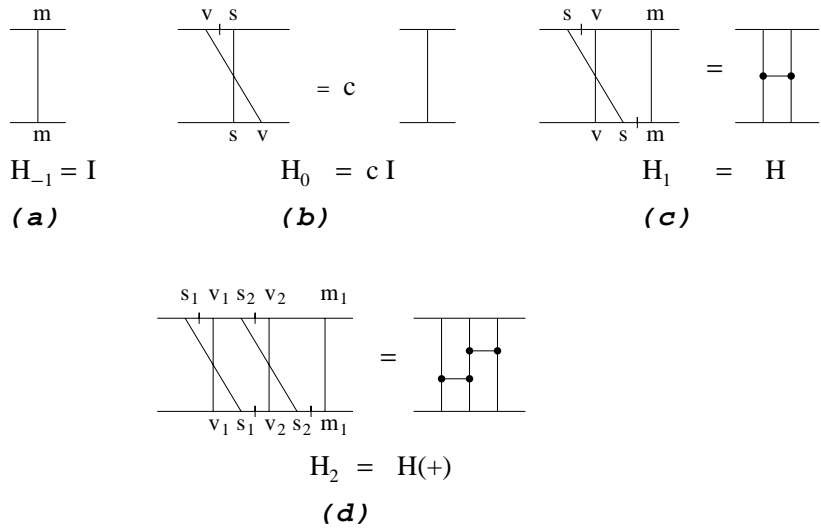}}
\nobreak
\vskip -1.5 cm\nobreak
\vskip .1 cm
\caption{Examples of how primitive color factors are obtained
from reduced diagrams.}
\end{figure}

The color factor of a reduced diagram may or may not be {\it connected}
after Fig.~3 is applied to get rid of the cuts, and after the upper and lower
quark lines are subsequently removed.
For example, those in Fig.~6 are connected and the one in Fig.~7 is not.
The set of connected diagrams in
${\cal C}$ will be denoted by ${\cal C}_0$.

To judge connectivity one may simply treat
the cut in SCC 
as a device to fuse together the pair of gluon lines it connects.
Cuts on the upper tree fuse upper ends of gluon lines, and these are always
the skeleton pairs. Cuts on
the lower tree fuse lower ends of the gluon lines. If such fusions
connect all gluon lines together then the diagram is connected.
For example,
the reduced diagram $[vs|m]$ in 6(c) is connected because the skeleton pair
$(vs)$ is fused at the upper end and the lines $(sm)$ are fused at the lower end.
Similarly, 6(d) is connected. However,
the diagram $[v_1m_1s_1|m_2v_2s_2m_3]$
in Fig.~7 is not, because fusion leaves it with four components consisting
of lines $(v_1s_1m_2)$, $(v_2s_2)$, $(m_1)$, and $(m_3)$.
\begin{figure}
\vskip -.5 cm
\centerline{\epsfxsize 3 truein \epsfbox {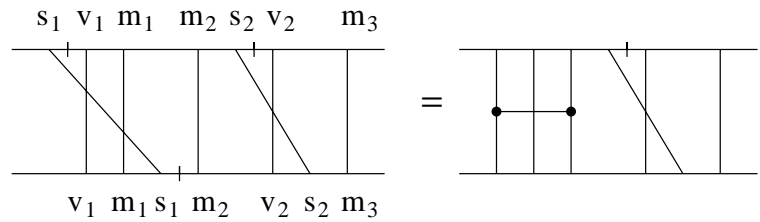}}
\nobreak
\vskip -2.5 cm\nobreak
\vskip .1 cm
\caption{An example of a disconnected reduced diagram and the corresponding
regge color factor.}
\end{figure}

The color factor of a disconnected reduced
diagram is given by the product of
the color factor of its connected components, according to the discussion of
Sec.~3 and App.~A.
\begin{figure}
\vskip .3 cm
\centerline{\epsfxsize 3 truein \epsfbox {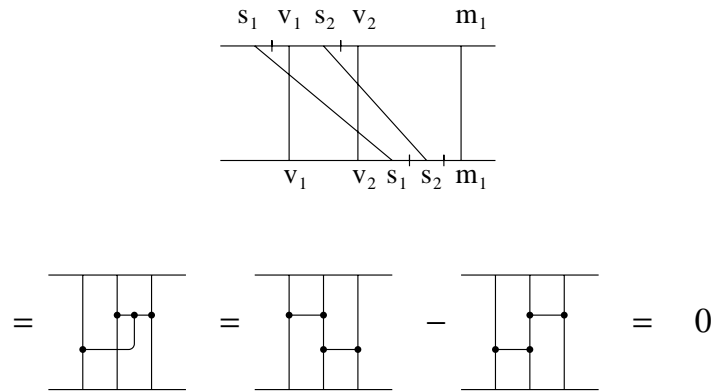}}
\nobreak
\vskip -1 cm\nobreak
\vskip .1 cm
\caption{An example of a connected reduced diagram that is not primitive.}
\end{figure}

\begin{figure}
\vskip -0 cm
\centerline{\epsfxsize 3 truein \epsfbox {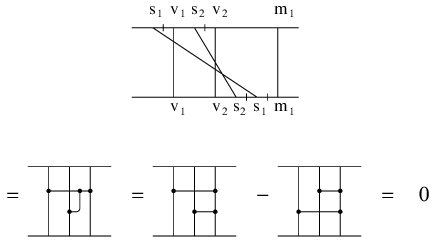}}
\nobreak
\vskip -2 cm\nobreak
\vskip .1 cm
\caption{Another example of a connected reduced diagram that is not
primitive.}
\end{figure}

The color factor of a connected component may or may not be primitive,
depending on whether it is one of those shown in
Fig.~1. The color factors in Fig.~6 are primitive,
but those in Figs.~8 and 9 are not.
However, as is shown in App.~A, those that are not all turn out to be zero,
so we need not worry about them at least for SCC diagrams.

The only primitive color factors encountered in SCC diagrams are
those shown in Fig.~6, and those similar to 6(d) but with $p>2$
skeleton crosses. They have $p$ horizontal lines, with the one
to the right always lying at a higher level.
These primitive color factors will be denoted as $H_p$, with $p=-1,
0,1,2,\cdots$ (see Fig.~6), and
the corresponding connected diagrams in ${\cal C}_0$
by $[H_p]$. It is convenient also to use the designation $I=H_{-1}$
and $H=H_1$ because of the similarity of these alphabets to the graphical
shape of the diagram. Note that $H_0=cH_{-1}=cI$, but all other primitive
color factors are independent.

In cut-tree notations, as members of ${\cal C}_0$, we have
\begin{eqnarray}
[H_{-1}]&=& [m] \nonumber\\
\ [H_{0}]&=& [vs] \nonumber\\
\ [H_{1}]&=& [vs|m] \nonumber\\
\ [H_{2}]&=& [v_1s_1|m_1v_2s_2|m_2] \nonumber\\
\ [H_{p}]&=& [v_1s_1|m_1v_2s_2|m_2\cdots v_ps_p|m_p]\ ,\quad(p\ge 1) \ \ \  
\end{eqnarray}

The reduced diagrams in ${\cal C}$ that give rise to
a regge color factor $\Phi=\prod_{p\ge -1}[(-)^pH_p]^{f_p}$ are simply
the {\it distinct} reduced diagrams obtained by interleaving $f_p$ copies of
$[H_p]$ together in all possible ways. In symbols,
\begin{eqnarray} 
\{\Phi\}=\{\prod_p(-)^{pf_p}H_p^{f_p}\}
=\{[H_{-1}];\cdots ;[H_0];\cdots ;[H_1];\cdots ;[H_2]
; \cdots  ; \cdots  \} 1/\prod_pf_p!\ ,
\end{eqnarray}
where the ellipses after each $[H_p]$ is an instruction to repeat the same
$[H_p]$ $f_p$ times, separated by semicolons. The sign $(-)^p$ associated
with $H_p$ comes about because of the minus sign associated with each cut.
The notation in (4.3)
for interleaving the cut trees in ${\cal C}_0$ is similar to the notation
$\{T_1;T_2;\cdots \}$ explained in Sec.~2 for interleaving
uncut trees $T_i$, but with two
differences. First,
lines separated by cuts are to be
thought of as being fused together by the cut, so lines from other cut trees
can never be inserted between them. Secondly, each cut diagram in
$\{[H_p];\cdots \}$ is going to occur $f_p!$ times because of the identical
nature of those diagrams. We allow only distinct diagrams in $\{\Phi\}$
so the division by $\prod_pf_p!$ in
(4.3) is a formal way of removing such redundancies.

The SCC diagrams in ${\cal S}$ that contain the reduced diagrams
in $\{\Phi\}$ will be denoted as $\{\Phi\}_{S}$. They can be obtained from
the cut trees in $\{\Phi\}$ by getting rid of their
cuts, which can be accomplished by moving the $s$-lines rightward in all
possible ways.
Instead of first interleaving the cut trees $[H_p]$ and then getting rid of the
cuts,
$\{\Phi\}_{S}$ can also be obtained by reversing the two operations, by
first getting rid of the cuts and then interleaving the uncut trees, in the
following way.

Start from $[H_p]\in{\cal C}_0$,
get rid of the cuts by moving the $s$-lines rightward, to construct all
$h_p^i\in{\cal S}\ (i=1,2,\cdots )$ that reduce to $[H_p]$.
In cases like 6(a) to 6(c) where there is only one tree for each $[H_p]$,
the degeneracy index $i=1$ will be omitted. This index is however needed in
other cases. For example, $[H_2]=[v_1s_1|m_1v_2s_2|m_2]$
in 6(d) gives rise to the uncut
trees $h_2^1=[v_1m_1s_1v_2m_2s_2]$, $h_2^2=[v_1m_1v_2s_1m_2s_2]$,
$h_2^3=[v_1m_1sv_2m_2s_1s_2]$, and $h_2^4=[v_1m_1v_2m_2s_2s_1]$.
The set of all $h_p^i\in{\cal S}$ for a fixed $p$ will be denoted by $\{H_p\}_{S}$.

The set of {\it distinct} SC or SCC diagrams obtained by interleaving $f_p$
trees in $\{H_p\}_{S}$ together is just $\{\Phi\}_{S}$.

\subsection{Factorization of sums of spacetime amplitudes}
We proceed to compute the sum of those spacetime amplitudes of all saturated SC
diagrams with a common regge color factor $\Phi=\prod_p ((-)^pH_p)^{f_p}$.
The relevant spacetime diagrams to be summed are those in the set
$\{\Phi\}_{S}$.

Using the factorization formula (2.5) on the {\it lower} tree, one gets
\begin{eqnarray} 
a\{\Phi\}_{S}=\sum_{{[T]\in\{\Phi\}}_{S}}a[T]=\prod_{p}{1\over f_p!}
\left(a\{H_p\}_{S}\right)^
{f_p}\ ,
\end{eqnarray}
where
\begin{eqnarray} 
{1\over f_p!}\left(a\{H_p\}_{S}\right)^{f_p}\equiv\sum {1\over m_i!}
\left(a[h_p^i]\right)^{m_i}\ ,
\end{eqnarray}
with the sum taken over all $m_i\ge 0$ subject to $\sum_im_i=f_p$.
Thus $a\{H_p\}_{S}=\sum_ia[h_p^i]$.
The factorials in the denominators of (4.5) arise because of the necessity
to keep only distinct diagrams in $\{\Phi\}_{S}$.

\begin{figure}
\vskip -0 cm
\centerline{\epsfxsize 3 truein \epsfbox {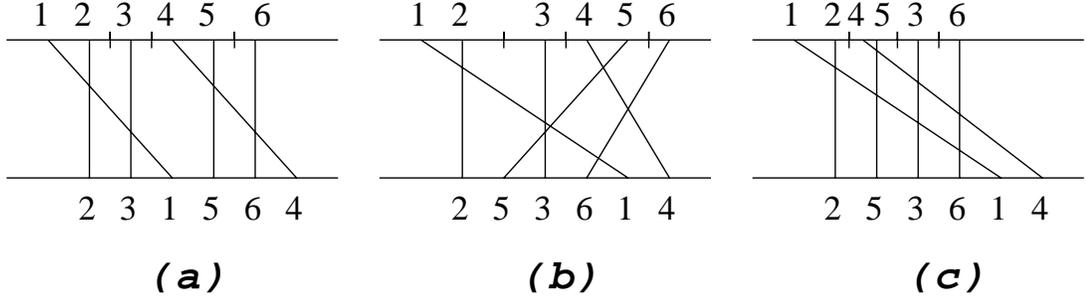}}
\nobreak
\vskip -2 cm\nobreak
\vskip .1 cm
\caption{Two SC diagrams in the set $\{231;564\}$. Diagram 10(b) is
identical to diagram 10(c).}
\end{figure}

The factorization (4.4) and (4.5) for the lower tree can be extended
to a factorization for the SC amplitudes. To do so we need to
use explicitly the cut property of the {\it upper} tree, that the
only uncut propagators are those inside the skeleton crosses.
To illustrate this point let us look at Fig.~10. Both the (lower) tree
[231564] in 10(a) and the tree [253614] in (10b) belong to the set
$\{231;564\}$,
but if we keep the upper ends of the gluon lines fixed
in 10(a) and 10(b), permuting the lower ends of the lines
to get from 10(a) to 10(b) does {\it not} change the SC diagram 10(a)
back to another SC diagram. 10(b), with lines 5 and 6 slanting the wrong way,
cannot be an SC diagram. However, by
making explicit use of the commuting properties of the amplitude for
the {\it upper tree}, $a[12|3|45|6]=a[12]a[3]a[45]a[6]=a[12|45|3|6]$,
10(b) can be redrawn as 10(c), which is an SC diagram. This can always
be done so factorization of the lower tree does lead to a factorization
of the sum of saturated SC amplitudes.

An SC diagram contains the lower tree, but it also contains
gluon propagators, quark propagators along the upper tree, vertex factors,
and loop integrations. In lightcone coordinates, $q_\pm=q^0\pm q^3$,
the measure of loop-integration is
\begin{eqnarray} 
{d^4q\over (2\pi)^4}={d^2q_\perp\over (2\pi)^2}{dq_+dq_-\over 8\pi^2}\ .
\end{eqnarray}
If the Dirac spinors are normalized to $\bar u u=1$, and a common factor is
taken out of the T-matrix amplitude ${\cal T}=-(s/2M^2){\cal A}$,
then each factorized amplitude $a\{H_p\}_{S}=\sum_ia[h_p^i]$ corresponds to a
saturated SC amplitude ${\cal A}\{H_p\}_{S}(\Delta)=\sum_i{\cal A}[h_p^i](\Delta)$,
where we have indicated explicitly the dependence on the momentum
transfer $\Delta$.
The product of two lower-tree amplitude $a\{H_a\}a\{H_b\}$
is turned into convolutions of two SC amplitudes:
\begin{eqnarray}
\left[ {\cal A}\{H_a\}_{S}*{\cal A}\{H_b\}_{S}\right](\Delta)
\equiv (-i)\int {d^2q_\perp\over
(2\pi)^2}\left[ {\cal A}\{H_a\}\right](\Delta-q_\perp)\left[ {\cal A}\{H_b\}\right](q_\perp)\ .
\end{eqnarray}
In obtaining (4.7), the identity
\begin{eqnarray} 
i\int {dq_+dq_-\over 8\pi^2}(-2\pi i)^2\delta(\sqrt{s}q_+)\delta
(\sqrt{s}q_-)(2s)=-i
\end{eqnarray}
has been used.

The sum of all saturated SC amplitudes with the regge color factor $\Phi$ is
then given by
\begin{eqnarray} 
\left[ {\cal A}\{\Phi\}_{S}\right](\Delta)=\prod_p
{1\over
f_p!}\left[ {\cal A}\{H_p\}_{S}\right]^{*f_p}(\Delta) \ .
\end{eqnarray}
All the products in (4.9) are meant to be convolutions. In particular,
$\left[ {\cal A}\{H_p\}_{S}\right]^{*f_p}$ is taken to mean $f_p$ convolutions of the same
amplitude. In impact-parameter space, such convolutions is replaced
by simple products.

\subsection{$O(g^6)$ results}

Let us now compare the general result of eq.~(4.9) with the $O(g^6)$ result
of Ref.~[7]. Except for the second-order tree diagram, they are shown
in Fig.~7 of Ref.~[7] as $B1_c, B2_c$, and $C15_c$ to $C20_c$.
The spacetime amplitudes are given in eq.~(6.1) of that reference
were expressed in terms of ${\cal M}=-{\cal A}/g^2$, hence
\begin{eqnarray}
{\cal A}({\bf G_1})&=&K_1-\ln s{c\over 2\pi}K_2\nonumber\\
{\cal A}({\bf G_2})&=&-{1\over 2} iK_2+i\ln s{c\over 2\pi}K_3\nonumber\\
{\cal A}({\bf G_3})&=&-i\ln s{1\over 2\pi}K_3\nonumber\\
{\cal A}({\bf G_4})&=&-{1\over 6}K_3\ ,
\end{eqnarray}
where the color factors ${\bf G_1, G_2, G_4, G_3}$
are the color factors $I, I^2, I^3, H$ in the present paper, and where
\begin{eqnarray} 
K_n(\Delta)=i^{n-1}(*K_1)^n,\quad K_1(\Delta)={g^2\over \Delta^2}\ .
\end{eqnarray}
The transverse functions $K_n$ is related to the ones used in Ref.~[7]
by $K_n=g^{2n}I_n$.

The general result according to (4.9) is
\begin{eqnarray} 
\left[ {\cal A}\{(-H)^aI^b\}_{S}\right](\Delta)
={1\over
a!b!}\left[ {\cal A}\{H_1\}_{S}\right]^{*a}*\left[ {\cal A}\{H_{-1}\}_S+
c{\cal A}\{H_0\}_{S}\right]^{*b}(\Delta)\ .
\end{eqnarray}
Substituting into (4.12) the explicit result obtained from eq.~(6.1)
of Ref.~[7],
\begin{eqnarray}
{\cal A}\{H_{-1}\}_{S}&=&{g^2\over \Delta^2}=K_1 \nonumber\\
{\cal A}\{H_0\}_{S}&=&-{g^2\over c}(B2_c)=-\ln s{1\over 2\pi}K_2\nonumber\\
{\cal A}\{H_1\}_{S}&=&-g^2(C20_c)=\ln s{i\over 2\pi}K_3\ ,
\end{eqnarray}
we get
\begin{eqnarray} 
\left[ {\cal A}\{(-H)^aI^b\}_{S}\right](\Delta)
={1\over
a!b!}\left[ \ln s{i\over 2\pi}K_3\right]^{*a}*\left[ K_1-
\ln s{c\over 2\pi}K_2\right]^{*b}(\Delta)\
.
\end{eqnarray}
The color factors ${\bf G_1,G_2,G_4}$ correspond to $(a,b)=(0,1), (0,2),
(0,3)$, and the color factor ${\bf G_3}$ corresponds to $(a,b)=(1,0)$
but with an extra minus sign. Expanding (4.14), keeping only leading-log
contributions and only up to $O(g^6)$, the result is the same as (4.10).

\subsection{Reggeized factorization}
The expression ${\cal A}\{H_{-1}\}_S+c{\cal A}\{H_0\}_S=K_1-(c/2\pi)\ln sK_2$
in (4.12) and (4.14) is the first two terms of the reggeized-gluon
propagator
\begin{eqnarray} 
R_1(\Delta, s)={g^2\over\Delta^2}\exp\left(-\bar\alpha(\Delta)\ln s\right)\ ,
\quad \bar\alpha(\Delta)={c\over 2\pi g^2}\Delta^2K_2(\Delta)\ .
\end{eqnarray}
The other terms come from $t$-channel-ladder and associated diagrams \cite{LN,CY,HC}
not considered here. The term ${\cal A}\{H_1\}_S$ in (4.12) is one of the many
terms contributing to the emission and reabsorption of an ordinary
gluon from a reggeized gluon, as indicated by the pattern H. Even to
$O(g^6)$, it receives contributions from other diagrams as well \cite{YJ}.
When all these are taken into account, it is known that
such emission and absorption
can be constructed from the Lipatov vertex \cite{LN} .

So the factorized results (4.9) and (4.12) are the beginning of
contributions that lead to {\it reggeized} factorization, but the
reggeization property cannot be seen fully without including other diagrams.
However, the reggeized nature of the color factors does seem to emerge
rather naturally.
\setcounter{equation}{0}
\section{Summary and outlook}

In this paper we initiated a leading-log investigation on sums of
Feynman diagrams contributing to multiple
reggeized-gluon exchanges. These diagrams are important because
they supply nonleading-log contributions to the $SU(3)_c$ gluon and Pomeron
amplitudes, thereby restoring unitarity. In any case
they supply the leading contributions in an $SU(N_c)$ color
theory in which the colliding beams carry large color so they must be taken
into account.

The central question studied in this paper is whether sums of Feynman
diagrams in the leading-log approximation will factorize into
multiple-reggeon-exchange
diagrams as depicted in Fig.~1. This `reggeized factorization
hypothesis' is nontrivial to prove for at least two reasons. First,
it is known that there are delicate cancellations in sums of Feynman
diagrams, so individual diagrams must be computed to subleading-log
accuracy to ensure a finite contribution to the sum. This is a very difficult
task for high order diagrams. Secondly, high-order Feynman diagrams are
very complicated, with lines criss-crossing in a complex pattern,
so it is far from obvious that they will sum up and factorize into neat
patterns as those displayed in Fig.~1. To date, factorization had been proved
completely only to the 6th order, and partially to the 8th and 10th orders,
by explicit calculations.

We prove in this paper the reggeized factorization hypothesis for
$s$-channel-ladder diagrams of any complexity. Both of the difficulties
mentioned above are solved by using instead the technique of
nonabelian cut diagrams discussed in a previous publication. These
cut diagrams are resummations of Feynman diagrams and are different from
the Cutkosky cut diagrams.

For other diagrams the validity of the reggeized factorization hypothesis
is still under investigation.

\section{Acknowlegements}

We thank Jean-Ren\'e Cudell and Omid Hamidi-Ravari for interesting
discussions. 
This research is supported in part by the by the Natural Science and
Engineering Research Council of Canada, and the Fonds pour la
Formation de Chercheurs et l'Aide \`a la Recherche of Qu\'ebec, and
YJF wishes to acknowledge the support of the Carl Reinhart Foundation.

\appendix
\setcounter{equation}{0}
\section{Color factors of nonabelian cut diagrams}
Color factor of nonabelian cut diagrams are calculated using the graphical
rules in Fig.~3. Some explicit examples are shown in Figs.~6--9.
In what follows, we shall discuss some of the general properties in the
leading-log approximation.

Fig.~3(c) can be used to get rid of cuts on the complementary cut diagrams.
As a result, diagrams with $m-1$ uncut propagators along the upper tree
has at most $m$ gluon lines attached to it. We say `at most', because
relations 3(b), 3(d), and 3(e) can sometimes be used to get rid of more lines.

Since cuts are made on the upper tree, the number of gluons $n$ attached to the
lower tree is often larger than the number $m$ attached to the upper tree.
However, by using Fig.~3 again to manipulate the lines attached to the lower
tree, at least in all cases encountered in Sec.~4, one can reduce the lines
attached to the lower tree to be $m$. Hence complementary
cut diagrams with $m-1$ uncut propagators along the upper tree contribute
color factors with $m$-reggeized-gluons, or less. It was then argued
in Sec.~3 of the text that we need not keep those with less than $m$ reggeons
in the leading-log approximation.

It is conceivable, for very complicated diagrams, that we cannot reduce $n$
to $m$ with the rules of Fig.~3 alone. The resulting color factor has $n\not=
m$, so it cannot possibly contribute if Fig.~1 is the final result. For that
reason we shall define the leading-log approximation to exclude all such
color factors that cannot be reduced to $n=m$.

Using 3(b) and 3(c) again, the positions of gluon lines attached to the
upper or the lower tree can be reversed; their difference being a diagram
with one less gluon line attached to the upper/lower tree,
and hence can be ignored in the leading-log
approximation. This is why primitive color factors can cross one another in any
way along the upper and the lower trees, yet giving exactly the same result
in the leading-log approximation.
\begin{figure}
\vskip -0 cm
\centerline{\epsfxsize 3 truein \epsfbox {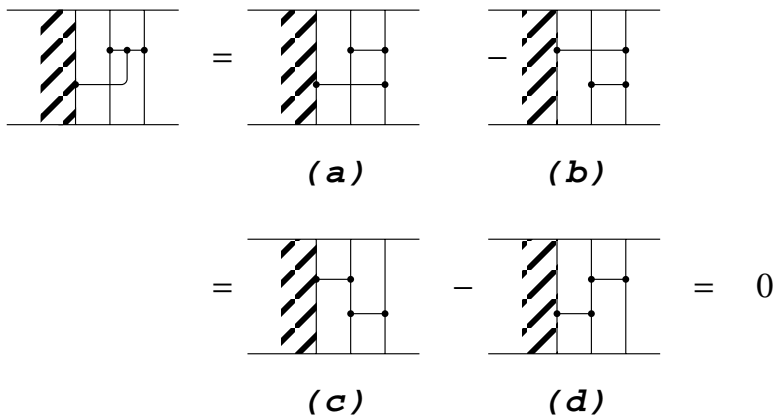}}
\nobreak
\vskip -2 cm\nobreak
\vskip .1 cm
\caption{Proof that an SCC diagram cannot yield a 
non-vanishing connected diagram that is not primitive. Within the leading-log
approximation, 11(c)=11(b) and 11(d)=11(a).}
\end{figure}

Finally, we want to prove that any color factor with an $s$-line climbing
onto the underside of a horizontal line, like those found in Figs.~8 and 9,
would be zero. The proof is shown in Fig.~11, where the shaded area can
contain a very complicated structure. Use 3(b) and 3(c) (for four gluon
lines) to move the point joining the bottom of the horizontal line to the
right, one gets 11(a) and 11(b). Moving that point to the left, one gets
11(c) and 11(d). Within the leading-log approximation, we can pull the
middle vertical line of 11(c) to the extreme right to get 11(b), hence
11(b)=11(c). Similarly 11(a)=11(d). Therefore 11(a)-11(b)=-[11(a)-11(b)]=0.

\section{Saturated ladder diagrams}
\setcounter{equation}{0}
We want to show in this Appendix that an SC diagram (see Sec.~4 for notation)
with two adjacent uncut propagators is unsaturated. By definition, a saturated
diagram of
$(2n)$th order and $m$rg exchange ($m-1$ cut lines) have a $g$ and $s$
dependence $g^{2m}(g^2\ln s)^{n-m}$. An unsaturated diagram is one with a
slower $s$ growth in comparison.

Ref.~[7] contains explicit calculations to $O(g^6)$. By examining Fig.~7
and eq.~(6.1) of that reference, it can
be seen that this assertion is valid to $O(g^6)$. If one now follows the
calculation of these examples with the method of Ref.~[5] and Appendix B
of Ref.~[7], one can see that these calculations can easily be generalized
to a multiloop situation as follows.

Consider an SC diagram with $n=\l-1$ gluon lines. Let 
$q_i=(q_{i+},q_{i-}, q_\perp)\ (1\le i\le \l)$ be the gluon momenta in 
the lightcone coordinates, and $q_{i-}\equiv x_i\sqrt{s}$. 

We shall follow Ref.~[5] by calculating the high energy behavior using
residue calculus and flow diagrams to carry out the `$+$' integrations..
For SC diagrams without adjacent uncut propagators, there is a
unique flow path for each diagram, and the poles for the `$+$' integration
can always be taken along the lower tree. The gluon propagators are
then $\sim 1/q_{i\perp}^2$, and considered to be $O(1)$.
This leaves the uncut propagators
along the upper tree to the `$-$' integration, each of which
contributes to a factor of $\ln s$ via `$-$' integration of the type 
$\int_{s^{-1}} dx_i/x_i$. Hence such diagrams have their full share of $\ln s$
factors and are saturated.

For diagrams with two adjacent uncut propagators, the flow path is never
unique: the flow direction along the boundary of the two adjacent uncut loops
cannot be determined. See Figs.~10.7 and 10.8 of Ref.~[5] for concrete
examples. As a result, at least one pole from the `$+$' integration
must {\it not} come from the lower tree. Explicit calculation then shows
that such diagrams are at least one $\ln s$ power down from the saturated
ones.

The origin of this reduction can be seen as follows.
The `$+$' momentum is inversely proportional
to the `$-$' momentum at the poles. Elsewhere the `$+$' momenta are determined
by momentum conservation. Now the `$-$' momentum flows predominantly
along the
lower tree, so if the pole is off it on a gluon line, the `$+$'
momentum flowing through that line would be relatively large. By momentum
conservation, there must
be a return flow passing through part of the lower tree and another gluon line,
and the Feynman propagators of these are large because of the large `$+$'
flow through them. This brings about at least two small factors $x_i$, 
overcompensating the large factor $1/x_i$ from the residue of the pole.
This costs at least a $\ln s$ factor to be lost from the `$-$' integration.
Hence the diagram is unsaturated.

\end{document}